\documentclass[preprintnumbers,amsmath,9pt,amssymb,floatfix,superscriptaddress,nofootinbib]{revtex4}
\usepackage{graphicx}
\usepackage{epsfig}
\usepackage{bm}
\usepackage{amsfonts}
\def\be {\begin{equation}}
\def\ee {\end{equation}}
\def\ba {\begin{eqnarray}}
\def\ea {\end{eqnarray}}

\begin{document}
\title{Power-Law and Logarithmic Entropy-Corrected Ricci Dark Energy in a Non-Flat FRW Universe with Viscous Interaction} 

 \author{\textbf{Antonio Pasqua}}\email{toto.pasqua@gmail.com} \affiliation{Department of Physics, University of Trieste, Trieste, Italy}

\begin{abstract}
\textbf{Abstract:} In this work, we consider the power-law corrected  and the logarithmic-corrected versions of the Holographic Dark Energy (HDE) model in a non-flat FRW Universe filled with a viscous Dark Energy (DE) interacting with Dark Matter (DM). We propose to replace the infrared cut-off with the inverse of the Ricci scalar curvature $R$. We obtain the equation of state (EoS) parameter $\omega_{\Lambda}$, the deceleration parameter $q$ and the evolution of energy density parameter $\Omega_{\Lambda} '$ in the presence of interaction between DE and DM for both corrections. We study the correspondence of the power-law entropy corrected Ricci dark energy (PLECRDE) and the logarithmic entropy corrected Ricci dark energy (LECRDE) models with the Generalized Chaplygin Gas (GCG), the Modified Variable Chaplygin Gas (MVCG), the New Modified Chaplygin Gas (NMCG),  the Yang-Mills (YM) and the Non Linear Electro-Dynamics (NLED) scalar field models.
%\textbf{Keywords:} Dark energy; cosmology; scalar fields; cosmic acceleration.
\end{abstract}
\maketitle

\newpage

\section{Introduction}

Observations from the Supernova Cosmology Project \cite{1,1-1}, WMAP \cite{2}, SDSS \cite{3}, and X-ray surveys \cite{4,4-1} indicate that the Universe’s expansion is accelerating. This unexpected behavior is attributed to dark energy (DE), an unknown component with negative pressure.

Several theoretical models have been proposed to describe DE, such as tachyon fields, K-essence, dilaton, quintessence, H-essence, and DBI-essence \cite{review,review-2}, though the simplest candidate remains the cosmological constant. Estimates from quantum field theory, using either Planck-scale or electroweak cutoffs, produce values far larger than observed—by factors of roughly $10^{123}$ and $10^{55}$—compared to $\Lambda/8\pi G \sim 10^{-47},\mathrm{GeV}^4$. The lack of a mechanism to explain such a small value is known as the cosmological constant problem \cite{sahni}.

A complete explanation of DE would require a consistent quantum gravity theory, which is still unavailable. Current frameworks like string theory and loop quantum gravity provide only approximate descriptions. String theory incorporates conjectures such as the AdS/CFT correspondence and the holographic principle, which implies that a system’s degrees of freedom scale with its surface area rather than its volume \cite{hooft}. Using this idea, the Holographic Dark Energy (HDE) model was introduced \cite{li,li-2} and has been extensively investigated in several works \cite{nood1,nood2,nood4,nood5,nood6,nood7}. In HDE, the quantum zero-point energy $\rho_\Lambda$ is constrained by black hole formation: the total energy in a region of size $L$ cannot exceed the mass of a black hole of the same size. Saturating this limit gives $\rho_\Lambda = 3 \alpha M_p^2 L^{-2}$ \cite{li,li-2}, where $\alpha$ is a constant and $M_p$ the reduced Planck mass.

The power-law correction to entropy which appear in
dealing with the entanglement of quantum fields in and
out the horizon is given by the following relation
 \cite{das,das-1,das-2}:
$$ S=c_0\Big( \frac{A}{a_1^2} \Big)[1+c_1f(A)],\ \ \ f(A)= \Big( \frac{A}{a_1^2} \Big)^{-\nu},$$
where the quantities $c_0$ and $c_1$ are two positive constant parameters of order unity, $a_1$ represents the ultraviolet cut-off at the horizon and $\nu$ is a fractional power which depends on the amount of mixing of ground and excited states. According to Das et al.~\cite{das}, when the horizon area is much larger than the cutoff scale ($A \gg a_1^2$), the correction term $f(A)$ becomes negligible, and the entanglement entropy of the mixed state gradually converges to the standard Bekenstein-Hawking entropy of the ground state.

Another form of entanglement entropy, which has proven useful, is given by \cite{power,power-1}:
\begin{equation}\label{1a}
S_h = \frac{A_h}{4} \Big( 1 - K_\alpha \, A_h^{1 - \frac{\alpha}{2}} \Big), \quad
K_\alpha = \frac{\alpha (4\pi)^{\frac{\alpha}{2}-1}}{(4-\alpha) r_c^{2-\alpha}},
\end{equation}
where $\alpha$ indicates a dimensionless parameter, $A_h = 4\pi R_h^2$ denotes the horizon area, $R_h$ is the horizon radius, and $r_c$ represents a crossover scale. To ensure that entropy remains well-defined, the parameter must satisfy $\alpha > 0$.

Quantum modifications to the standard entropy-area relation induce corresponding curvature corrections in the Einstein-Hilbert action, and conversely. Another commonly considered correction is the logarithmic form of entropy \cite{jamil5,jamil5-1,jamil5-2,jamil5-3}:
\begin{equation}\label{1b}
S_h = \frac{A_h}{4} + \beta \log\Big(\frac{A_h}{4}\Big) + \gamma,
\end{equation}
where $\beta$ and $\gamma$ are constants. Such corrections naturally arise in the context of loop quantum gravity, reflecting both thermal equilibrium fluctuations and quantum fluctuations of black hole entropy.

It is worth emphasizing that various other generalizations of entropy have also been proposed in \cite{odi1}.

Building upon the quantum-corrected forms of entropy discussed above, recently proposed models of Holographic Dark Energy (HDE) incorporate these entropy corrections. Using the power-law corrected entropy from (\ref{1a}), the corresponding energy density can be expressed as \cite{power,power-1}:
\begin{equation}\label{1c}
\rho_\Lambda = 3 \alpha M_p^2 L^{-2} - \beta M_p^2 L^{-\gamma}.
\end{equation}
In a similar manner, the logarithmic correction in Eq. (\ref{1b}) leads to the following form for the energy density \cite{wei-1,wei-2,wei-3}:
\begin{equation}\label{1d}
\rho_\Lambda = 3 \alpha M_p^2 L^{-2} + \gamma_1 L^{-4} \log(M_p^2 L^2) + \gamma_2 L^{-4}.
\end{equation}

Gao et al.~\cite{gao} observed that choosing the future event horizon as the infrared cutoff in holographic dark energy (HDE) models introduces a causality issue. To address this, they suggested using the Ricci scalar of the FRW metric as an alternative cut-off, $L = R^{-1/2}$. This choice not only alleviates the causality problem but also provides a natural solution to the coincidence problem, giving rise to the so-called \textit{Ricci Dark Energy} (RDE) model. According to Gao et al., a value of $\alpha \simeq 0.46$ reproduces the observed dark energy density and the current equation of state accurately. 

Furthermore, the RDE framework shows consistency with various observational datasets, including supernovae, cosmic microwave background radiation, baryon acoustic oscillations, gas mass fractions in galaxy clusters, the Hubble parameter evolution, and the growth rate of cosmic structures \cite{pavon}.

Motivated by the work of Gao et al.~\cite{gao}, we extend the Ricci dark energy model by incorporating power-law entropy corrections, leading to the energy density
\begin{equation}\label{5-1}
    \rho_{\Lambda} = 3 \alpha M_p^2 R - \beta M_p^2 R^{\gamma/2},
\end{equation}
where $R$ denotes the Ricci scalar, defined as
\begin{equation}
R = 6 \Big( \dot{H} + 2 H^2 + \frac{k}{a^2} \Big).
\end{equation}
Here, $H = \dot{a}(t)/a(t)$ is the Hubble parameter, $\dot{H}$ is its time derivative, $a(t)$ is the scale factor, and $k$ represents the spatial curvature with dimensions of $[\text{length}]^{-2}$. The values $k=-1, 0, +1$ correspond to an open, flat, or closed FRW Universe, respectively.

In a similar fashion, the energy density of the logarithmic entropy-corrected Ricci dark energy can be expressed as:
\begin{equation}\label{5}
    \rho_{\Lambda} = 3 \alpha M_p^2 R + \gamma_1 R^2 \log\Big(\frac{M_p^2}{R}\Big) + \gamma_2 R^2.
\end{equation}
The Ricci DE model has been widely and well studied in many different papers \cite{riccide1,riccide2,riccide3,riccide4,riccide5,riccide6,riccide7,riccide8,riccide9,riccide10,riccide12,riccide13,riccide14,riccide15,riccide16,riccide17,riccide18}

The paper is structured as follows. In Section~2, we outline the physical framework and derive expressions for the equation of state (EoS) parameter $\omega_{\Lambda}$, the deceleration parameter $q$, and the evolutionary fomr of the EoS parameter $\Omega_{\Lambda}'$ for our models in a non-flat FRW Universe. Section~3 is devoted to establishing a correspondence between our models and various scalar field theories, including the Generalized Chaplygin Gas (GCG), the Modified Variable Chaplygin Gas (MVCG), the New Modified Chaplygin Gas (NMCG), the Yang-Mills (YM) and the Non Linear Electro-Dynamics (NLED) scalar field models. Finally, Section~4 presents the main conclusions of this work.

\section{INTERACTING MODEL IN A NON-FLAT Universe}

We consider the background spacetime to be described by a spatially homogeneous and isotropic Friedmann–Robertson–Walker (FRW) metric, whose line element is given by
\begin{equation}\label{6}
    ds^2 = -dt^2 + a^2(t) \left[ \frac{dr^2}{1 - k r^2} + r^2 (d\theta^2 + \sin^2\theta\, d\varphi^2) \right],
\end{equation}
where $t$ denotes the cosmic time, $r$ is the radial coordinate, and $(\theta, \varphi)$ are the usual angular coordinates.

The corresponding Friedmann equation in this background reads
\begin{equation}\label{7}
    H^2 + \frac{k}{a^2} = \frac{1}{3 M_p^2} \left( \rho_\Lambda + \rho_m \right),
\end{equation}
where $\rho_\Lambda$ and $\rho_m$ represent the energy densities of dark energy and dark matter, respectively.

We introduce the dimensionless density parameters for matter, curvature, and dark energy as follows:
\begin{eqnarray}
    \Omega_m &=& \frac{\rho_m}{\rho_{cr}} = \frac{\rho_m}{3 M_p^2 H^2}, \label{8}\\
    \Omega_k &=& \frac{\rho_k}{\rho_{cr}} = \frac{k}{H^2 a^2}, \label{9}\\
    \Omega_\Lambda &=& \frac{\rho_\Lambda}{\rho_{cr}} = \frac{\rho_\Lambda}{3 M_p^2 H^2}, \label{10}
\end{eqnarray}
where the critical density is defined by $\rho_{cr} = 3 M_p^2 H^2$. The parameter $\Omega_k$ measures the relative contribution of the spatial curvature to the total density. Current observational constraints indicate a small but nonzero value, $\Omega_k \simeq 0.02$ \cite{spergel}.

Using Eqs.~(\ref{8})–(\ref{10}), the Friedmann equation given in Eq. (\ref{7}) can be expressed in terms of the density parameters as
\begin{eqnarray}
1 + \Omega_k = \Omega_m + \Omega_\Lambda. \label{11}
\end{eqnarray}
In order to ensure the validity of the Bianchi identity, i.e. the local conservation of the energy–momentum tensor $\nabla_\mu T^{\mu\nu}=0$, 
the total energy density $\rho_{tot} = \rho_\Lambda + \rho_m$ must obey the conservation equation
\begin{eqnarray}
\dot{\rho} + 3H(1+\omega)\rho = 0, \label{12}
\end{eqnarray}
where the parameters are defined as follows:
\begin{eqnarray}
\omega = \frac{p}{\rho}, \qquad
\rho = \rho_m + \rho_\Lambda, \qquad
p = \bar{p}_\Lambda = p_\Lambda - 3\varepsilon H .
\end{eqnarray}
Here, $\omega$ is the effective equation-of-state (EoS) parameter, $\rho$ denotes the total energy density, and $p$ is the effective pressure of dark energy. 
The quantity $\xi$ represents the bulk viscosity coefficient \cite{ren,ren-2,tanmoy}. Since dark matter is pressureless, one has $p_m = 0$. 
Following \cite{karami}, the bulk viscosity can be parameterized as $\xi = \varepsilon \rho_\Lambda H^{-1}$, where $\varepsilon$ is a constant. 
With this assumption, the total pressure becomes
\begin{eqnarray}
p = (\omega_\Lambda - 3\varepsilon)\rho_\Lambda,
\end{eqnarray}
where $\omega_\Lambda = p_{tot}/\rho_{tot}$ denotes the EoS parameter of viscous dark energy.

If we further allow for a direct interaction between dark matter and dark energy, their energy densities evolve according to separate balance equations \cite{sheykhi}:
\begin{eqnarray}
\dot{\rho}_m + 3H \rho_m &=& Q, \label{13}\\
\dot{\rho}_\Lambda + 3H\rho_\Lambda (1+\omega_\Lambda) &=& -Q + 9\varepsilon H \rho_\Lambda, \label{14}
\end{eqnarray}
where $Q$ is the interaction term, which may in general depend on $H$, $\rho_m$ and $\rho_\Lambda$, i.e. $Q(H\rho_m, H\rho_\Lambda)$. 
The most commonly adopted phenomenological form is \cite{umar,umar-1,umar-2,umar-3,umar-4,umar-5}
\begin{eqnarray}
Q = 3b^2 H (\rho_m + \rho_\Lambda), \label{15}
\end{eqnarray}
with $b^2$ a dimensionless coupling parameter. Several different functional forms of $Q$ have been investigated in the literature \cite{rashid,rashid-1}, 
and observational limits on $b^2$ have been derived \cite{b2}. High-resolution $N$-body simulations further indicate that an interaction between dark matter and dark energy 
may significantly affect the internal properties of nonlinear cosmic structures, such as the concentration–mass relation \cite{salu}. 
Since the growth of dark matter perturbations is particularly sensitive to such an interaction, the value of the coupling parameter $b^2$ 
can strongly influence the cosmic evolution \cite{maru,maru-2}. Ideally, a fundamental expression for $Q$ should arise from a consistent theory of quantum gravity,  or alternatively from a reconstruction scheme based on observational data such as SNIa \cite{sn,sn-2}. Despite the lack of a fully established theoretical basis,  interacting dark energy models provide an excellent fit to present observations \cite{wang}.

We now turn our attention to the determination of the equation of state (EoS) parameter $\omega_{\Lambda}$ for the logarithmic entropy-corrected Ricci dark energy (LECRDE) and the power-law entropy-corrected Ricci dark energy (PLECRDE) models. \\
The explicit form of $\omega_{\Lambda}$ can be obtained by combining the energy density expressions of these models with the conservation equations discussed above. This derivation will allow us to investigate how the entropy corrections affect the dynamical properties of dark energy within the interacting viscous framework.

Using the Friedamm equation (\ref{7}), the Ricci scalar $R$ can be written as:
\begin{eqnarray}
    R=6\Big(  \dot{H} + H^2 + \frac{\rho_m+\rho_{\Lambda}}{3M_p^2} \Big). \label{17}
\end{eqnarray}
From the Friedmann equation (\ref{7}), we also derive
\begin{eqnarray}
    \dot{H}&=&\frac{k}{a^2}-\frac{1}{2M_p^2}\Big( \rho + \bar{p}  \Big), \nonumber \\
   % = \frac{k}{a^2} - \frac{1}{2M_p^2} \Big( \rho_m + \rho_{\Lambda} + \omega_{\Lambda}\rho_{\Lambda}-3\xi H \Big)= \nonumber \\
    %= \frac{k}{a^2} - \frac{1}{2M_p^2} \Big( \rho_m + \Big( 1+\omega_{\Lambda} \Big)\rho_{\Lambda}-3\varepsilon \rho_{\Lambda}  \Big) \nonumber \\
     &=& \frac{k}{a^2} - \frac{1}{2M_p^2}\Big[ \rho_m + ( 1+\omega_{\Lambda}-3\varepsilon )\rho_{\Lambda} \Big].\label{dotH}
\end{eqnarray}
Adding Eqs.~(\ref{7}) and (\ref{dotH}), we obtain
\begin{eqnarray}    
\dot{H}+H^2 &=& \frac{1}{3M_p^2}\Big(\rho_m+\rho_{\Lambda}\Big)
-\frac{\rho_m}{2M_p^2}
-\frac{1}{2M_p^2}\Big( 1+\omega_{\Lambda} - 3\varepsilon  \Big)\rho_{\Lambda}.   
\label{jamil20}
\end{eqnarray}
Therefore, the Ricci scalar $R$ given in Eq.~(\ref{17}) can be rewritten as
\begin{eqnarray}    
R &=& \frac{\rho_m}{M_p^2}
+\frac{3\rho_{\Lambda}}{M_p^2}\Big(\tfrac{1}{3}-\omega_{\Lambda}+3\varepsilon \Big). 
\label{jamil22}
\end{eqnarray}
From Eq.~(\ref{jamil22}) we can easily derive the expression for the EoS parameter $\omega_{\Lambda}$:
\begin{eqnarray}  
\omega_{\Lambda}
&=& 3\varepsilon-\frac{RM_p^2}{3\rho_{\Lambda}}
+\frac{\rho_{\Lambda}+\rho_{m}}{3\rho_{\Lambda}} \nonumber \\
&=& 3\varepsilon-\frac{RM_p^2}{3\rho_{\Lambda}}
+\frac{\Omega_{\Lambda}+\Omega_{m}}{3\Omega_{\Lambda}},
\label{21}
\end{eqnarray}
where we have used the relation 
\(
\frac{\rho_{\Lambda}+\rho_{m}}{3\rho_{\Lambda}} 
= \frac{\Omega_{\Lambda}+\Omega_{m}}{3\Omega_{\Lambda}}
\).

Substituting Eq.~(\ref{5-1}) into Eq.~(\ref{21}) and using Eq.~(\ref{11}), we obtain for the PLECRDE model:
\begin{eqnarray}
\omega_{\Lambda,{\rm pl}}
&=& 3\varepsilon 
-\frac{1}{3\Big(3\alpha -\beta R^{\gamma /2-1}\Big)}
+\frac{\Omega_m+\Omega_{\Lambda}}{3\Omega_{\Lambda}} \nonumber \\
&=& 3\varepsilon 
-\frac{1}{3\Big(3\alpha -\beta R^{\gamma /2-1}\Big)}
+\frac{1+\Omega_k}{3\Omega_{\Lambda}}. 
\label{rplechde}
\end{eqnarray}
Similarly, substituting in Eq.~(\ref{21}) the expression of the energy density $\rho_{\Lambda}$ given in Eq.~(\ref{5}) and using Eq.~(\ref{11}), we obtain for the LECRDE model:
\begin{eqnarray}
\omega_{\Lambda,{\rm log}}
&=& 3\varepsilon 
-\frac{M_p^2/3}{3\alpha M_p^2+\gamma_1R\log\!\Big(\tfrac{M_p^2}{R}\Big)+\gamma_2R}
+\frac{\Omega_m+\Omega_{\Lambda}}{3\Omega_{\Lambda}} \nonumber \\
&=& 3\varepsilon 
-\frac{M_p^2/3}{3\alpha M_p^2+\gamma_1R\log\!\Big(\tfrac{M_p^2}{R}\Big)+\gamma_2R}
+\frac{1+\Omega_k}{3\Omega_{\Lambda}}.    
\label{22}
\end{eqnarray}

We emphasize that in both cases we have made use of Eq.~(\ref{11}).

We now proceed to derive the expression governing the evolution of the energy density parameter $\Omega_{\Lambda}$.  
From Eq.~(\ref{14}), the equation of state (EoS) parameter $\omega_{\Lambda}$ can be written as
\begin{eqnarray}
    \omega_{\Lambda} = -1 - \frac{\dot{\rho}_{\Lambda}}{3H\rho_{\Lambda}} - \frac{Q}{3H\rho_{\Lambda}} + 9\varepsilon. \label{23}
\end{eqnarray}

To determine $\dot{\rho}_{\Lambda}$, we start from the continuity equation for the dark energy component, which includes both an interaction term $Q$ and a bulk-viscosity contribution:
\begin{equation}
\dot{\rho}_\Lambda + 3H(1 + \omega_\Lambda)\rho_\Lambda = -Q + 9H^2 \varepsilon \rho_\Lambda.
\end{equation}

The interaction between dark energy and dark matter is assumed to take the form
\begin{equation}
Q = 3Hb^2(\rho_m + \rho_\Lambda).
\end{equation}
Substituting this expression into the continuity equation gives
\begin{equation}
\dot{\rho}_\Lambda = -3H(1 + \omega_\Lambda)\rho_\Lambda - 3Hb^2(\rho_m + \rho_\Lambda) + 9H^2 \varepsilon \rho_\Lambda.
\end{equation}

Next, we employ the expression of the EoS parameter $\omega_\Lambda$ obtained from the scalar curvature $R$:
\begin{equation}
\omega_\Lambda = 3\varepsilon - \frac{R M_p^2}{3\rho_\Lambda} + \frac{\rho_\Lambda + \rho_m}{3\rho_\Lambda}.
\end{equation}
Thus, we have
\begin{align}
(1 + \omega_\Lambda)\rho_\Lambda &= \left( 1 + 3\varepsilon - \frac{R M_p^2}{3\rho_\Lambda} + \frac{\rho_\Lambda + \rho_m}{3\rho_\Lambda} \right)\rho_\Lambda \nonumber \\
&= \rho_\Lambda + 3\varepsilon \rho_\Lambda - \frac{R M_p^2}{3} + \frac{\rho_\Lambda + \rho_m}{3}.
\end{align}

Substituting this result into the expression for $\dot{\rho}_\Lambda$ yields
\begin{align}
\dot{\rho}_\Lambda &= -3H\left( \rho_\Lambda + 3\varepsilon \rho_\Lambda - \frac{R M_p^2}{3} + \frac{\rho_\Lambda + \rho_m}{3} \right) - 3H b^2(\rho_m + \rho_\Lambda) + 9H^2 \varepsilon \rho_\Lambda \nonumber \\
&= 3H\left[ -\rho_\Lambda - (\rho_\Lambda + \rho_m)\left( b^2 + \tfrac{1}{3} \right) + \frac{R M_p^2}{3} + 3\varepsilon \rho_\Lambda \right], \label{picu1}
\end{align}
which constitutes the final expression for $\dot{\rho}_{\Lambda}$.  

Dividing by the critical density $\rho_c = 3H^2 M_p^2$, Eq.~(\ref{picu1}) becomes
\begin{eqnarray}
    \frac{\dot{\rho}_{\Lambda}}{\rho_c}=3H\Big[ -\Omega_{\Lambda}-\big(1 + \Omega_k\big)\Big(b^2+\tfrac{1}{3} \Big) + \frac{R}{9H^2}+3\varepsilon\Omega_{\Lambda} \Big] = \dot{\Omega_{\Lambda}}+2\Omega_{\Lambda}\left(\frac{\dot{H}}{H}\right).  \label{25}
\end{eqnarray}

Using Eq.~(\ref{17}), the term $\tfrac{R}{9H^2}$ can be recast as
\begin{eqnarray}
    \frac{R}{9H^2} = \frac{2}{3}\left( \frac{\dot{H}}{H^2} + 2 + \Omega_k \right). \label{26}
\end{eqnarray}

Substituting Eq.~(\ref{26}) into Eq.~(\ref{25}), we obtain the time evolution of $\Omega_\Lambda$:
\begin{eqnarray}
\dot{\Omega}_{\Lambda} = 2\frac{\dot{H}}{H}\Big(1-\Omega_{\Lambda}\Big)+3H\Big[-\Omega_{\Lambda}-\big(1 + \Omega_k\big)\Big(b^2-\tfrac{1}{3}\Big) + \tfrac{2}{3}+ 3\varepsilon\Omega_{\Lambda}\Big]. \label{27}
\end{eqnarray}

Since $\Omega_{\Lambda}' = \tfrac{d\Omega_{\Lambda}}{dx}= \tfrac{1}{H}\dot{\Omega}_{\Lambda}$ with $x = \ln a$, Eq.~(\ref{27}) becomes
\begin{eqnarray}
H \, \Omega_{\Lambda}' = 2H'\Big(1-\Omega_{\Lambda}\Big)+3H\Big[-\Omega_{\Lambda}-\big(1 + \Omega_k\big)\Big(b^2-\tfrac{1}{3}\Big) + \tfrac{2}{3}+ 3\varepsilon\Omega_{\Lambda}\Big], \label{28}
\end{eqnarray}
which simplifies to
\begin{eqnarray}
    \Omega_{\Lambda}'=\frac{2}{H}\Big(1-\Omega_{\Lambda}\Big)+3\Big[-\Omega_{\Lambda}-\big(1 + \Omega_k\big)\Big(b^2-\tfrac{1}{3}\Big) + \tfrac{2}{3}+3\varepsilon\Omega_{\Lambda}\Big]. \label{29}
\end{eqnarray}
In Eq.~(\ref{29}) we used the fact that
\begin{eqnarray}
    H'=\frac{a'}{a}=1. \label{30}
\end{eqnarray}

\subsection*{Deceleration parameter}
For completeness, let us also compute the deceleration parameter $q$, defined as
\begin{eqnarray}
	q=-\frac{\ddot{a}a}{\dot{a}^2}=  -\frac{\ddot{a}}{aH^2}  =-1-\frac{\dot{H}}{H^2}. \label{33}
\end{eqnarray}

Taking the derivative of the Friedmann equation (\ref{7}) with respect to cosmic time and making use of Eqs.~(\ref{11}), (\ref{13}) and (\ref{14}), the deceleration parameter can be recast as
\begin{eqnarray}
    q=\frac{1}{2}\Big[1 + \Omega_k + 3\Omega_{\Lambda} \omega_{\Lambda} \Big]. \label{32}
\end{eqnarray}

Using the expression of $\omega_{\Lambda, pl}$ given in Eq.~(\ref{rplechde}), the deceleration parameter for the PLECRDE model can be written as:
\begin{eqnarray}
    q_{pl}=1-\frac{1}{2}\left(\frac{\Omega _{\Lambda}}{3\alpha -\beta R^{\gamma /2-1}}\right)+ \Omega _k +\frac{9\varepsilon \Omega_{\Lambda}}{2}.
\end{eqnarray}

Similarly, substituting the expression of the EoS parameter $\omega_{\Lambda, log}$ from Eq.~(\ref{22}), the deceleration parameter for the LECRDE model reads
\begin{eqnarray}
    q_{log}=1-\frac{1}{2}\left[\frac{M_p^2\Omega _{\Lambda}}{3\alpha M_p^2+\gamma_1R \log\!\big(M_p^2R^{-1}\big)+\gamma_2R}\right]+ \Omega _k +\frac{9\varepsilon \Omega_{\Lambda}}{2}. \label{jamil26}
\end{eqnarray}

\subsection*{Limiting case: flat dark-energy–dominated Universe}
We now focus on the limiting case of a flat, dark-energy–dominated Universe within the HDE model.  
For the LECRDE case, this corresponds to $\gamma_1=\gamma_2=0$, $\Omega_{\Lambda}=1$, and $\Omega_k=\Omega_m=0$.  
For the PLECRDE case, the conditions are $\beta=0$, $\Omega_{\Lambda}=1$, and $\Omega_k=\Omega_m=0$.  

In both cases, the dark energy density given in Eqs.~(\ref{5}) and (\ref{5-1}) reduces to
\begin{eqnarray}
\rho_{\Lambda}=3\alpha M_{p}^2 R. \label{34}
\end{eqnarray}

From the Friedmann equation (\ref{7}), the Hubble parameter $H$ and the Ricci scalar curvature $R$ are obtained as
\begin{eqnarray}
H=\frac{6\alpha}{12\alpha-1}\Big(\frac{1}{t}\Big), \label{35}
\end{eqnarray}
\begin{eqnarray}
R=\frac{36\alpha}{(12\alpha-1)^2}\Big(\frac{1}{t^2}\Big). \label{36}
\end{eqnarray}

Therefore, the dark energy density can be expressed as
\begin{eqnarray}
\rho_{\Lambda}= \frac{108M_p^2\alpha^2}{(12\alpha-1)^2}\Big(\frac{1}{t^2}\Big). \label{34-2}
\end{eqnarray}

Finally, integrating the Hubble parameter $H(t)$, the scale factor evolves as
\begin{equation}
a(t) = a_0 \, t^{\frac{6\alpha}{12\alpha - 1}}.
\end{equation}

From Eq.~(\ref{21}), we obtain
\begin{eqnarray}  
\omega_{\Lambda} &=& 3\varepsilon - \frac{R M_p^2}{3\rho_{\Lambda}} + \frac{1}{3}.
\end{eqnarray}
Using the expressions of $\rho_{\Lambda}$ and $H$ derived above, this reduces to
\begin{eqnarray}
    \omega_{\Lambda} = \frac{1}{3} - \frac{1}{9\alpha} + 3\varepsilon. \label{36-1}
\end{eqnarray}

Accordingly, the deceleration parameter takes the form
\begin{eqnarray}
    q = 1 - \frac{1}{6\alpha} + \frac{9\varepsilon}{2}. \label{36-2}
\end{eqnarray}

In the limiting case $\varepsilon = 0$, Eqs.~(\ref{36-1}) and (\ref{36-2}) reduce to
\begin{eqnarray}
\omega_{\Lambda} = \frac{1}{3} - \frac{1}{9\alpha}, \label{LEoS}
\end{eqnarray}
\begin{eqnarray}
q = 1 - \frac{1}{6\alpha}, \label{Lq}
\end{eqnarray}
which coincide with the results obtained in the absence of bulk viscosity.  

From Eq.~(\ref{LEoS}), we see that in this limit the EoS parameter of dark energy becomes constant.  
For $\alpha < 1/12$, one finds $\omega_{\Lambda} < -1$, meaning that the phantom divide can be crossed.  
Since the Ricci scalar diverges at $\alpha = 1/12$, this value must be excluded.  
From Eq.~(\ref{Lq}), we further deduce that cosmic acceleration starts at $\alpha \leq 1/6$, which also marks the onset of the quintessence regime ($\omega_{\Lambda} \leq -1/3$).

\section{CORRESPONDENCE BETWEEN THE R-PLECHDE MODEL  AND SCALAR FIELDS}
In this Section, we establish a correspondence between the R-PLECHDE model and a set of scalar field models, namely: the Generalized Chaplygin Gas (GCG), the Modified Chaplygin Gas (MCG), the Modified Variable Chaplygin Gas (MVCG), the New Modified Chaplygin Gas (NMCG), the Yang--Mills (YM), and the Non--Linear Electrodynamics (NLED) models.  

This choice is motivated by the fact that scalar field models are widely regarded as an effective description of dark energy.  
To this end, we first compare the energy density of the R-PLECHDE model with that of the corresponding scalar field model.  
Subsequently, we equate the equation of state (EoS) parameter of each scalar field model under consideration with the EoS parameter of the R-PLECHDE model.

\subsection{The Generalized Chaplygin Gas (GCG) Scalar Field Model}
We begin by establishing a correspondence between the R-PLECHDE model and the first scalar field model considered, namely the Generalized Chaplygin Gas (GCG).  

In a seminal work, Kamenshchik \emph{et al.}~\cite{gcg1} introduced the homogeneous Chaplygin Gas (CG) model, based on a single fluid obeying the equation of state (EoS) 
\begin{eqnarray}
    p = -\frac{A_0}{\rho},
\end{eqnarray}
where $p$ and $\rho$ denote the pressure and energy density of the fluid, respectively, and $A_0$ is a positive constant parameter.  
A subsequent generalization of this model led to the Generalized Chaplygin Gas (GCG).  

The GCG model is particularly appealing since it interpolates between a dust-dominated phase and a late-time accelerated expansion, thereby providing a good fit to observational cosmological data~\cite{gcg6}. Its EoS is defined as~\cite{gcg9,gcg16,gcg17}
\begin{eqnarray}
    p_D = -\frac{D}{\rho_{\Lambda}^{\theta}}, \label{gcg1}
\end{eqnarray}
where $D$ and $\theta$ are free constant parameters, with $D > 0$ and $0 < \theta < 1$.  
The standard Chaplygin Gas is recovered in the special case $\theta = 1$.  

Historically, the EoS in Eq.~(\ref{gcg1}) with $\theta = 1$ was first studied by Chaplygin in 1904 in the context of adiabatic processes~\cite{gcg1}.  
The case $\theta \neq 1$ was later analyzed in~\cite{gcg3}. The possibility that a Chaplygin gas cosmological model could unify dark energy (DE) and dark matter (DM) was originally proposed for $\theta = 1$ in~\cite{gcg4,gcg5}, and subsequently generalized to $\theta \neq 1$ in~\cite{gcg3}.  

Gorini \emph{et al.}~\cite{gcg15} showed that the matter power spectrum is compatible with observations only if $\theta < 10^{-5}$, in which case the GCG model becomes practically indistinguishable from the standard $\Lambda$CDM cosmology.  
In~\cite{gcg7}, Chaplygin inflation was studied within the framework of Loop Quantum Cosmology, where it was shown that the model parameters are consistent with the five-year WMAP data.  

The evolution of the GCG energy density is given by
\begin{eqnarray}
    \rho_{\Lambda} = \left[ D + \frac{B}{a^{3(\theta+1)}} \right]^{\frac{1}{\theta+1}}, \label{gcg2}
\end{eqnarray}
where $B$ is an integration constant.

In principle, Eq.~(\ref{gcg2}) allows a wide range of positive values for the parameter $\theta$.  
However, it must ensure that the sound speed
\begin{eqnarray}
c_s^2 = \frac{D \theta}{\rho^{\theta + 1}}
\end{eqnarray}
does not exceed the speed of light $c$.  
Furthermore, as pointed out by Bento \emph{et al.}~\cite{gcg3}, only for $0 < \theta < 1$ does the analysis of energy density fluctuations have physical significance.

We now reconstruct the potential and dynamics of this scalar field model within the framework of the DE model under study.  
The energy density $\rho_{\Lambda}$ and the pressure $p_D$ of a homogeneous and time-dependent scalar field $\phi$ are given by
\begin{eqnarray}
\rho_{\Lambda} &=& \frac{1}{2}\dot{\phi}^2 + V(\phi), \label{gcg3}\\
p_D &=& \frac{1}{2}\dot{\phi}^2 - V(\phi). \label{gcg4}
\end{eqnarray}
From Eqs.~(\ref{gcg3}) and (\ref{gcg4}), the EoS parameter reads
\begin{eqnarray}
\omega_{\Lambda} = \frac{\frac{1}{2}\dot{\phi}^2 - V(\phi)}{\frac{1}{2}\dot{\phi}^2 + V(\phi)}
= \frac{\dot{\phi}^2 - 2V(\phi)}{\dot{\phi}^2 + 2V(\phi)}. \label{gcg5}
\end{eqnarray}

Adding Eqs.~(\ref{gcg3}) and (\ref{gcg4}) gives the kinetic term
\begin{eqnarray}
\dot{\phi}^2 = \rho_{\Lambda} + p_D. \label{gcg6-1}
\end{eqnarray}
Using the definition of $p_D$ in Eq.~(\ref{gcg1}), we have
\begin{eqnarray}
\dot{\phi}^2 = \rho_{\Lambda} - \frac{D}{\rho_{\Lambda}^{\theta}} = \rho_{\Lambda} \left(1 - \frac{D}{\rho_{\Lambda}^{\theta+1}}\right). \label{gcg6-11}
\end{eqnarray}

Considering $\dot{\phi} = H \phi'$ and the expression of $\rho_{\Lambda}$ in Eq.~(\ref{gcg2}), the evolutionary form of $\phi$ is
\begin{eqnarray}
\phi = \int_{a_0}^a \sqrt{3 \omega_{\Lambda}} 
\sqrt{1 - \frac{D}{D + \frac{B}{a^{3(\theta+1)}}}} \frac{da}{a}. \label{gcg6-1newint}
\end{eqnarray}
For a flat, dark-energy-dominated Universe, the solution is
\begin{eqnarray}
\phi(a) &=& \frac{2\sqrt{3}}{3(1+\theta)} 
\left[ \log\left(a^{\frac{3}{2}(1+\theta)}\right) - \log\left(B + \sqrt{B(B + a^{3(1+\theta)} D)}\right) \right]. \label{murano4}
\end{eqnarray}

Subtracting Eqs.~(\ref{gcg3}) and (\ref{gcg4}) gives the scalar potential
\begin{eqnarray}
V(\phi) = \frac{1}{2} (\rho_{\Lambda} - p_D) = \frac{\rho_{\Lambda}}{2} \left(1 + \frac{D}{\rho_{\Lambda}^{\theta+1}}\right). \label{gcg7-1}
\end{eqnarray}
Using Eq.~(\ref{gcg2}), this becomes
\begin{eqnarray}
V(\phi) = \frac{1}{2} \left[ D + \frac{B}{a^{3(\theta+1)}} \right]^{\frac{1}{\theta+1}} 
+ \frac{1}{2}\cdot \frac{D}{\left[ D + \frac{B}{a^{3(\theta+1)}} \right]^{\frac{\theta}{\theta+1}}}. \label{gcg7-1new}
\end{eqnarray}

The parameters $D$ and $B$ in terms of cosmological quantities are obtained from
\begin{eqnarray}
\omega_{\Lambda} = -\frac{D}{\rho_{\Lambda}^{\theta+1}} \quad \Rightarrow \quad
D = -\omega_{\Lambda} \rho_{\Lambda}^{\theta+1}, \label{gcg9}
\end{eqnarray}
and
\begin{eqnarray}
B = a^{3(\theta+1)} \left( \rho_{\Lambda}^{\theta+1} - D \right) = \left(a^3 \rho_{\Lambda} \right)^{\theta+1} (1 + \omega_{\Lambda}). \label{BBB}
\end{eqnarray}

Using the R-PLECHDE EoS parameter $\omega_{\Lambda,pl}$, we find
\begin{eqnarray}
D_{pl} &=& \frac{\rho_{\Lambda,pl}^{\theta+1}}{3} \left[-3\varepsilon + \frac{1}{3(3\alpha - \beta R^{\gamma/2-1})} - \frac{1+\Omega_k}{3\Omega_{\Lambda}} \right], \label{murano6} \\
B_{pl} &=& \left(a^3 \rho_{\Lambda,pl}\right)^{\theta+1} \left[1 + 3\varepsilon - \frac{1}{3(3\alpha - \beta R^{\gamma/2-1})} + \frac{1+\Omega_k}{3\Omega_{\Lambda}} \right]. \label{murano7}
\end{eqnarray}

Similarly, for the R-LECHDE model:
\begin{eqnarray}
D_{log} &=& \frac{\rho_{\Lambda,log}^{\theta+1}}{3} \left[-3\varepsilon + \frac{M_p^2/3}{3\alpha M_p^2 + \gamma_1 R \log(M_p^2 R^{-1}) + \gamma_2 R} - \frac{1+\Omega_k}{3\Omega_{\Lambda}} \right], \\
B_{log} &=& \left(a^3 \rho_{\Lambda,log}\right)^{\theta+1} \left[1 + 3\varepsilon - \frac{M_p^2/3}{3\alpha M_p^2 + \gamma_1 R \log(M_p^2 R^{-1}) + \gamma_2 R} + \frac{1+\Omega_k}{3\Omega_{\Lambda}} \right].
\end{eqnarray}

Finally, in the flat, dark-energy-dominated limit, using Eq.~(\ref{36-1}), we obtain
\begin{eqnarray}
D_{Dark} &=& -\left(\frac{1}{3} - \frac{1}{9\alpha} + 3\varepsilon\right) \left[ \frac{108 M_p^2 \alpha^2}{(12\alpha - 1)^2} \frac{1}{t^2} \right]^{\theta+1}, \label{murano10} \\
B_{Dark} &=& \left(1 + \frac{1}{3} - \frac{1}{9\alpha} + 3\varepsilon\right) \left[ \frac{108 M_p^2 \alpha^2}{(12\alpha - 1)^2} t^{-\frac{6\alpha-2}{12\alpha-1}} \right]^{\theta+1}.
\end{eqnarray}

In the limiting case $\varepsilon = 0$, these reduce to
\begin{eqnarray}
D_{Dark,lim} &=& -\left(\frac{1}{3} - \frac{1}{9\alpha}\right) \left[ \frac{108 M_p^2 \alpha^2}{(12\alpha - 1)^2} \frac{1}{t^2} \right]^{\theta+1}, \\
B_{Dark,lim} &=& \left(1 + \frac{1}{3} - \frac{1}{9\alpha}\right) \left[ \frac{108 M_p^2 \alpha^2}{(12\alpha - 1)^2} t^{-\frac{6\alpha-2}{12\alpha-1}} \right]^{\theta+1}.
\end{eqnarray}

Furthermore, using the general definition of the DE EoS parameter $\omega_{\Lambda}$, we can rewrite Eqs.~(\ref{gcg6-1}) and (\ref{gcg7-1}) as
\begin{eqnarray}
\dot{\phi}^2 &=& \left(1 + \omega_{\Lambda}\right)\rho_{\Lambda}, \label{gcg6mura1}\\
V(\phi) &=& \frac{1}{2} \left(1 - \omega_{\Lambda}\right)\rho_{\Lambda}. \label{gcg7mura2}
\end{eqnarray}

Using the R-PLECHDE EoS parameter $\omega_{\Lambda,pl}$, the kinetic and potential terms become
\begin{eqnarray}
\dot{\phi}^2_{pl} &=& \rho_{\Lambda,pl} \left[ 1 + 3\varepsilon - \frac{1}{3(3\alpha - \beta R^{\gamma/2-1})} + \frac{1+\Omega_k}{3\Omega_{\Lambda}} \right], \label{gcg6mura1model1}\\
V_{pl}(\phi) &=& \frac{\rho_{\Lambda,pl}}{2} \left[ 1 - 3\varepsilon + \frac{1}{3(3\alpha - \beta R^{\gamma/2-1})} - \frac{1+\Omega_k}{3\Omega_{\Lambda}} \right]. \label{gcg7mura2model1}
\end{eqnarray}

The evolutionary form of the scalar field is obtained by integrating Eq.~(\ref{gcg6mura1model1}) with respect to the scale factor $a(t)$:
\begin{eqnarray}
\phi_{pl}(a) - \phi_{pl}(a_0) &=& \int_{a_0}^{a} \sqrt{3 \Omega_{\Lambda,pl}} 
\left[ 1 + 3\varepsilon - \frac{1}{3(3\alpha - \beta R^{\gamma/2-1})} + \frac{1+\Omega_k}{3\Omega_{\Lambda}} \right]^{1/2} \frac{da}{a}. \label{gcg19}
\end{eqnarray}

Similarly, for the R-LECHDE model, we have
\begin{eqnarray}
\dot{\phi}^2_{log} &=& \rho_{\Lambda,log} \left[ 1 + 3\varepsilon - \frac{M_p^2/3}{3\alpha M_p^2 + \gamma_1 R \log(M_p^2 R^{-1}) + \gamma_2 R} + \frac{1+\Omega_k}{3\Omega_{\Lambda}} \right], \\
V_{log}(\phi) &=& \frac{\rho_{\Lambda,log}}{2} \left[ 1 - 3\varepsilon + \frac{M_p^2/3}{3\alpha M_p^2 + \gamma_1 R \log(M_p^2 R^{-1}) + \gamma_2 R} - \frac{1+\Omega_k}{3\Omega_{\Lambda}} \right],
\end{eqnarray}
with the evolutionary scalar field
\begin{eqnarray}
\phi_{log}(a) - \phi_{log}(a_0) &=& \int_{a_0}^{a} \sqrt{3 \Omega_{\Lambda,log}} 
\left[ 1 + 3\varepsilon - \frac{M_p^2/3}{3\alpha M_p^2 + \gamma_1 R \log(M_p^2 R^{-1}) + \gamma_2 R} + \frac{1+\Omega_k}{3\Omega_{\Lambda}} \right]^{1/2} \frac{da}{a}.
\end{eqnarray}

In the limiting case of a flat, dark-energy-dominated Universe, i.e., $\omega_{\Lambda,pl} = \Omega_{\Lambda,log} = 1$, $\Omega_k = \Omega_m = 0$, the scalar field and potential reduce to
\begin{eqnarray}
\dot{\phi}^2(t) &=& \frac{108 M_p^2 \alpha^2}{(12\alpha - 1)^2} \left( \frac{4}{3} - \frac{1}{9\alpha} + 3\varepsilon \right) \frac{1}{t^2}, \label{murano12} \\
V(t) &=& \frac{54 M_p^2 \alpha^2}{(12\alpha - 1)^2} \left( \frac{2}{3} + \frac{1}{9\alpha} - 3\varepsilon \right) \frac{1}{t^2}. \label{murano13gcg}
\end{eqnarray}

Integrating $\dot{\phi}$, we get
\begin{eqnarray}
\phi(t) &=& \sqrt{\frac{108 M_p^2 \alpha^2}{(12\alpha - 1)^2} \left( \frac{4}{3} - \frac{1}{9\alpha} + 3\varepsilon \right)} \ln t + \text{constant}.
\end{eqnarray}

For $\varepsilon = 0$, these reduce to
\begin{eqnarray}
\phi(t) &=& \sqrt{\frac{12 M_p^2 \alpha}{12\alpha - 1}} \ln t + \text{constant}, \\
V(t) &=& \frac{6 M_p^2 \alpha (6\alpha + 1)}{(12\alpha - 1)^2} \frac{1}{t^2}. \label{murano13gcg_lim}
\end{eqnarray}

\subsection{The Modified Variable Chaplygin Gas (MVCG) Scalar Field Model}
We now consider the Modified Variable Chaplygin Gas (MVCG) model. Guo and Zhang \cite{mvcg1} recently proposed the Variable Chaplygin Gas (VCG) model, whose equation of state (EoS) is given by  

\begin{equation}  
p_D = - \frac{B}{\rho_{\Lambda}}, \label{murano16}  
\end{equation}  

where \(B\) is assumed to be a function of the scale factor \(a(t)\), i.e., \(B = B(a(t))\). This choice is physically reasonable, as it can be related to a scalar potential if the Chaplygin Gas is interpreted via a Born-Infeld scalar field \cite{mvcg2}. For simplicity, in the following we will omit the explicit temporal dependence of the scale factor. The VCG model has been recently studied in \cite{mvcg3,mvcg4}.  

Debnath \cite{mvcg5} proposed the EoS of the Modified Variable Chaplygin Gas (MVCG) in the form  

\begin{equation}  
p_D = A \rho_{\Lambda} - \frac{B(a)}{\rho_{\Lambda}^{\theta}}, \label{murano17}  
\end{equation}  

where \(A\), \(\theta\), and \(B(a)\) are parameters. In this paper, we adopt the specific choice \(B(a) = B_0 a^{-\delta_1}\), so that the pressure of the MVCG model can be written as  

\begin{equation}  
p_D = A \rho_{\Lambda} - \frac{B_0 a^{-\delta_1}}{\rho_{\Lambda}^{\theta}}. \label{murano18}  
\end{equation}  

Here, \(A\), \(B_0\), and \(\delta_1\) are positive constants, with \(B_0\) representing the present-day value of \(B\) and \(\delta_1\) the exponent of the scale factor. The parameter \(\theta\) is typically taken in the range \(0 \leq \theta \leq 1\).  

In the limiting case \(B_0 = 0\), Eq. (\ref{murano18}) reduces to a barotropic EoS, \(p = A \rho\), which can describe various types of matter: \(A = -1\) corresponds to a cosmological constant; \(A = -2/3\) to domain walls; \(A = -1/3\) to cosmic strings; \(A = 0\) to dust or pressureless matter; \(A = 1/3\) to relativistic gas; \(A = 2/3\) to a perfect gas; and \(A = 1\) to ultra-stiff matter.  

If we instead take \(B = B_0\) constant (i.e., \(\delta_1 = 0\)), Eq. (\ref{murano18}) reduces to the EoS of the original Modified Chaplygin Gas model. In general, the MVCG scenario interpolates between a radiation-dominated phase (\(A = 1/3\)) and a quintessence-dominated phase characterized by negative pressure. In the specific limit \(A = 0\) and \(\theta = 1\), the usual Chaplygin Gas is recovered. Recent analyses using the latest Supernovae data indicate that models with \(\theta > 1\) are also viable \cite{mvcg8}. Moreover, when \(A = 0\), Eq. (\ref{murano18}) describes a fluid with negative pressure, characteristic of the quintessence regime.

This modified form of the Chaplygin Gas also has a phenomenological motivation, as it can account for the flat rotation curves of galaxies \cite{mvcg9}. The galactic rotational velocity \(V_c\) is related to the MVCG parameter \(A\) through the relation
\begin{equation}
V_c = \sqrt{2A},
\end{equation}
while the density parameter \(\rho\) is related to the radial size of the galaxy \(r\) by
\begin{equation}
\rho = \frac{A}{2 \pi G r^2}.
\end{equation}
At high densities, the first term in the MVCG EoS dominates, producing flat rotation curves that are consistent with current observations. The parameter \(A\) varies from galaxy to galaxy due to differences in \(V_c\).

The energy density \(\rho_{\Lambda}\) of the MVCG model is given by
\begin{equation}
\rho_{\Lambda} = \left\{ \frac{3(\theta +1)B_0}{\left[ 3(\theta +1)(A+1) - \delta_1 \right]} \frac{1}{a^{\delta_1}} - \frac{C}{a^{3(\theta +1)(A+1)}} \right\}^{\frac{1}{1+\theta}}, \label{murano19}
\end{equation}
where \(C\) is a positive integration constant, and we require \(3(\theta +1)(A+1) > \delta_1\) to ensure that the first term is positive. The parameter \(\delta_1\) must also be positive; otherwise, as \(a \to \infty\), the energy density \(\rho_{\Lambda}\) would diverge, which is not compatible with an expanding Universe.

We now reconstruct the potential and dynamics of the scalar field. For this purpose, we consider a time-dependent scalar field \(\phi(t)\) with potential \(V(\phi)\), which are directly related to the energy density and pressure of the MVCG as
\begin{align}
\rho_{\Lambda} &= \frac{1}{2} \dot{\phi}^2 + V(\phi), \label{murano20} \\
p_D &= \frac{1}{2} \dot{\phi}^2 - V(\phi). \label{murano21}
\end{align}
Since the kinetic term is positive, the MVCG behaves as a quintessence-type scalar field.

The deceleration parameter \(q\) is defined as
\begin{equation}
q = -\frac{\ddot{a}}{aH^2}, \label{murano22}
\end{equation}
where \(a\) is the scale factor and \(H\) the Hubble parameter. For an accelerating Universe, we require \(q < 0\), which implies \(\ddot{a} > 0\), since \(a > 0\) and \(H^2 > 0\). This condition leads to
\begin{equation}
\left[ \frac{2(1+\theta) - \delta_1}{3(1+\theta)(1+A) - \delta_1} \right] a^{3(1+\theta)(1+A) - \delta_1} > \frac{C(1+3A)}{3B_0}. \label{murano23}
\end{equation}

Equation \eqref{murano23} requires that \(\delta_1 < 2(1+\theta)\). Since \(0 \leq \theta \leq 1\), we obtain the range
\begin{equation}
0 < \delta_1 < 4.
\end{equation}

This condition shows that for small values of the scale factor \(a\), the Universe is decelerating, whereas for large values of \(a\), it accelerates. The transition occurs at
\begin{equation}
a = \left\{ \frac{C(1+3A) [3(1+\theta)(1+A) - \delta_1]}{3B_0 [2(1+\theta) - \delta_1]} \right\}^{\frac{1}{3(1+\theta)(1+A) - \delta_1}}. \label{murano24}
\end{equation}

For small scale factor \(a(t)\), the energy density behaves as
\begin{equation}
\rho \simeq \frac{C^{\frac{1}{1+\theta}}}{a^{3(1+A)}}, \label{murano25}
\end{equation}
which corresponds to a Universe dominated by a barotropic fluid with equation of state \(p = A \rho\).

For large values of the scale factor, the energy density behaves as
\begin{equation}
\rho \simeq \left[ \frac{3(1+\theta) B_0}{3(1+\theta)(1+A) - \delta_1} \right]^{\frac{1}{1+\theta}} a^{-\frac{\theta}{1+\theta}}, \label{murano26}
\end{equation}
which corresponds to a quintessence-like equation of state
\begin{equation}
p = \left( \frac{\delta_1}{3(1+\theta)} - 1 \right) \rho. \label{murano27}
\end{equation}

In the limiting case \(\delta_1 = 0\), Eq. \eqref{murano27} reduces to the original modified Chaplygin gas scenario. In general, the variable modified Chaplygin gas interpolates between a radiation-dominated phase (\(A = 1/3\)) and a quintessence-dominated phase described by a constant EoS 
\(\gamma = -1 + \frac{\delta_1}{3(1+\theta)} < -\frac{1}{3}\).

The positivity of the energy density given in Eq. \eqref{murano19} imposes a lower bound on the scale factor:
\begin{equation}
a(t) > \left\{ -\frac{C \left[ 3(\theta+1)(A+1) - \delta_1 \right]}{3(\theta+1) B_0} \right\}^{\frac{1}{3(\theta+1)(A+1) - \delta_1}} \equiv a_{\min}(t). \label{murano29}
\end{equation}

The scalar field \(\phi(t)\) associated with the MVCG can be reconstructed from the relations
\begin{align}
\dot{\phi}^2 &= \rho_{\Lambda} + p_D, \label{murano30} \\
V(\phi) &= \frac{\rho_{\Lambda} - p_D}{2}, \label{murano31}
\end{align}
where \(\rho_\Lambda\) and \(p_D\) are given by Eqs. \eqref{murano19} and \eqref{murano18}, respectively. Substituting them, we obtain
\begin{align}
\dot{\phi}^2 &= (1+A) \left\{ \frac{3(\theta+1) B_0}{3(\theta+1)(A+1) - \delta_1} \frac{1}{a^{\delta_1}} - \frac{C}{a^{3(\theta+1)(A+1)}} \right\}^{\frac{1}{1+\theta}} \nonumber \\
&\quad - \frac{B_0 a^{-\delta_1}}{\left\{ \frac{3(\theta+1) B_0}{3(\theta+1)(A+1) - \delta_1} \frac{1}{a^{\delta_1}} - \frac{C}{a^{3(\theta+1)(A+1)}} \right\}^{\frac{\theta}{1+\theta}}}. \label{murano32}
\end{align}

Using \(\dot{\phi} = H \frac{d\phi}{da} a\), the scalar field can be expressed as
\begin{align}
\phi &= \int_{t_0}^{t} \sqrt{1+A} \left\{ \frac{3(\theta+1) B_0}{3(\theta+1)(A+1) - \delta_1} \frac{1}{a^{\delta_1}} - \frac{C}{a^{3(\theta+1)(A+1)}} \right\}^{\frac{1}{2(1+\theta)}} dt \nonumber \\
&\quad - \int_{t_0}^{t} \frac{\sqrt{B_0 a^{-\delta_1}}}{\left\{ \frac{3(\theta+1) B_0}{3(\theta+1)(A+1) - \delta_1} \frac{1}{a^{\delta_1}} - \frac{C}{a^{3(\theta+1)(A+1)}} \right\}^{\frac{\theta}{2(1+\theta)}}} dt. \label{murano33}
\end{align}

For a flat Universe (\(k=0\)), the Friedmann equation \eqref{7} gives the explicit dependence of \(t\) on \(a(t)\) as
\begin{equation}
t = K a^{\frac{\delta_1}{2(1+\theta)}} \, _2F_1 \left[ \frac{1}{2(1+\theta)}, -z, 1-z, -\frac{C}{K} a^{-\frac{\delta_1}{2(1+\theta) z}} \right], \label{murano34}
\end{equation}
where
\begin{align}
K &= \frac{2}{\delta_1} \left[ (1+\theta)^{\theta} \sqrt{\frac{\delta_1}{6 B_0 z}} \right]^{\frac{1}{1+\theta}}, \label{murano35} \\
z &= \frac{\delta_1}{2(1+\theta) [3(1+A)(1+\theta) - \delta_1]}. \label{murano36}
\end{align}

Here, \(_2F_1\) is the Gaussian hypergeometric function of the second kind.

For a flat Universe, considering the expressions of $t$ given in Eq. \eqref{murano34}, the scalar field $\phi$ can be written as
\begin{align}
\phi &= \frac{\sqrt{1+A}}{3(1+A)(1+\theta) - \delta_1} \Bigg\{ 
2 \log \left( \sqrt{u+x} + \sqrt{u+y} \right) \nonumber \\
&\quad - \sqrt{\frac{y}{x}} \log \left[ \frac{ \left( \sqrt{x(u+x)} + \sqrt{y(u+y)} \right)^2 }{ x^{3/2} \sqrt{y} u } \right] \Bigg\}, \label{murano37}
\end{align}
where
\begin{align}
x &= \frac{\delta_1}{1+A}, \label{murano38} \\
y &= 3(1+\theta), \label{murano39} \\
u &= \left( \frac{\delta_1 C}{B_0} \right) a^{\delta_1 \left( 1 - \frac{y}{x} \right)}. \label{murano40}
\end{align}

The scalar potential is obtained from Eq. \eqref{murano31} as
\begin{align}
V(\phi) &= \frac{1-A}{2} \left\{ \frac{3(\theta+1)B_0}{3(\theta+1)(A+1)-\delta_1} \frac{1}{a^{\delta_1}} - \frac{C}{a^{3(\theta+1)(A+1)}} \right\}^{\frac{1}{1+\theta}} \nonumber \\
&\quad + \frac{B_0 a^{-\delta_1}}{2 \left\{ \frac{3(\theta+1)B_0}{3(\theta+1)(A+1)-\delta_1} \frac{1}{a^{\delta_1}} - \frac{C}{a^{3(\theta+1)(A+1)}} \right\}^{\frac{\theta}{1+\theta}}}. \label{murano40}
\end{align}

Dividing $p_D$ by $\rho_\Lambda$, the EoS parameter reads
\begin{equation}
\omega_\Lambda = A - \frac{B_0 a^{-\delta_1}}{\rho_\Lambda^{\theta+1}}, \label{murano41}
\end{equation}
which gives
\begin{equation}
B_0 = a^{\delta_1} (A - \omega_\Lambda) \rho_\Lambda^{\theta+1}. \label{murano42}
\end{equation}

From the expression of $\rho_\Lambda$ we also get
\begin{equation}
C = \left[ \frac{3(\theta+1)B_0}{3(\theta+1)(A+1) - \delta_1} \frac{1}{a^{\delta_1}} - \rho_\Lambda^{1+\theta} \right] a^{-3(\theta+1)(A+1)}, \label{murano43}
\end{equation}
and inserting Eq. \eqref{murano42} into Eq. \eqref{murano43}, we obtain
\begin{equation}
C = \left( \rho_\Lambda a^{-3(A+1)} \right)^{\theta+1} \left[ \frac{3(\theta+1)(A-\omega_\Lambda)}{3(\theta+1)(A+1) - \delta_1} - 1 \right]. \label{murano44}
\end{equation}

For the R-PLECHDE model, substituting $\omega_\Lambda$ yields
\begin{align}
B_{0,pl} &= a^{\delta_1} \left[ A - 3\varepsilon + \frac{1}{3(3\alpha - \beta R^{\gamma/2-1})} - \frac{1+\Omega_k}{3\Omega_\Lambda} \right] \rho_{\Lambda,pl}^{\theta+1}, \label{murano45} \\
C_{pl} &= \left( \rho_{\Lambda,pl} a^{-3(A+1)} \right)^{\theta+1} \left\{ \frac{3(\theta+1) \left[ A - 3\varepsilon + \frac{1}{3(3\alpha - \beta R^{\gamma/2-1})} - \frac{1+\Omega_k}{3\Omega_\Lambda} \right]}{3(\theta+1)(A+1) - \delta_1} - 1 \right\}. \label{murano46}
\end{align}

Similarly, for the R-LECHDE model:
\begin{align}
B_{0,log} &= a^{\delta_1} \left[ A - 3\varepsilon + \frac{M_p^2/3}{3\alpha M_p^2 + \gamma_1 R \log(M_p^2 R^{-1}) + \gamma_2 R} - \frac{1+\Omega_k}{3\Omega_\Lambda} \right] \rho_{\Lambda,log}^{\theta+1}, \\
C_{log} &= \left( \rho_{\Lambda,log} a^{-3(A+1)} \right)^{\theta+1} \left\{ \frac{3(\theta+1) \left[ A - 3\varepsilon + \frac{M_p^2/3}{3\alpha M_p^2 + \gamma_1 R \log(M_p^2 R^{-1}) + \gamma_2 R} - \frac{1+\Omega_k}{3\Omega_\Lambda} \right]}{3(\theta+1)(A+1) - \delta_1} - 1 \right\}.
\end{align}

In the flat Dark Dominated Universe, we have
\begin{align}
B_{0,Dark} &= \left( A - \frac{1}{3} + \frac{1}{9\alpha} - 3\varepsilon \right) \left[ \frac{108 M_p^2 \alpha^2}{(12\alpha-1)^2} \right]^{\theta+1} t^{\frac{6\alpha \delta_1}{12\alpha-1} - 2(\theta+1)}, \label{murano49} \\
C_{Dark} &= \left[ \frac{108 M_p^2 \alpha^2}{(12\alpha-1)^2} t^{\frac{-6\alpha(3A+7)}{12\alpha-1}} \right]^{\theta+1} \left[ \frac{3(\theta+1)\left( A - \frac{1}{3} + \frac{1}{9\alpha} - 3\varepsilon \right)}{3(\theta+1)(A+1)-\delta_1} - 1 \right]. \label{murano50}
\end{align}

For \(\varepsilon = 0\), we obtain the limiting case:
\begin{align}
B_{0,Dark,lim} &= \left( A - \frac{1}{3} + \frac{1}{9\alpha} \right) \left[ \frac{108 M_p^2 \alpha^2}{(12\alpha-1)^2} \right]^{\theta+1} t^{\frac{6\alpha \delta_1}{12\alpha-1} - 2(\theta+1)}, \\
C_{Dark,lim} &= \left[ \frac{108 M_p^2 \alpha^2}{(12\alpha-1)^2} t^{\frac{-6\alpha(3A+7)}{12\alpha-1}} \right]^{\theta+1} \left[ \frac{3(\theta+1)\left( A - \frac{1}{3} + \frac{1}{9\alpha} \right)}{3(\theta+1)(A+1)-\delta_1} - 1 \right].
\end{align}

Using the general definition of the EoS parameter of DE, $\omega_\Lambda$, we can rewrite Eqs. \eqref{gcg6-1} and \eqref{gcg7-1} as
\begin{align}
\dot{\phi}^2 &= (1+\omega_\Lambda)\rho_\Lambda, \label{gcg6mura1} \\
V(\phi) &= \frac{1}{2}(1-\omega_\Lambda)\rho_\Lambda. \label{gcg7mura2}
\end{align}

For the R-PLECHDE model, inserting $\omega_{\Lambda,pl}$  gives
\begin{align}
\dot{\phi}^2_{pl} &= \rho_{\Lambda,pl} \left[ 1 + 3\varepsilon - \frac{1}{3(3\alpha - \beta R^{\gamma/2-1})} + \frac{1+\Omega_k}{3\Omega_\Lambda} \right], \label{gcg6mura1model1} \\
V_{pl}(\phi) &= \frac{\rho_{\Lambda,pl}}{2} \left[ 1 - 3\varepsilon + \frac{1}{3(3\alpha - \beta R^{\gamma/2-1})} - \frac{1+\Omega_k}{3\Omega_\Lambda} \right]. \label{gcg7mura2model1}
\end{align}

The evolutionary form of the scalar field is obtained integrating Eq. \eqref{gcg6mura1model1} with respect to the scale factor $a(t)$:
\begin{align}
\phi(a)_{pl} - \phi(a_0)_{pl} &= \int_{a_0}^{a} \sqrt{3\Omega_{\Lambda,pl}} \, 
\left[ 1 + 3\varepsilon - \frac{1}{3(3\alpha - \beta R^{\gamma/2-1})} + \frac{1+\Omega_k}{3\Omega_\Lambda} \right]^{1/2} \frac{da}{a}. \label{gcg19pl}
\end{align}

Similarly, for the R-LECHDE model:
\begin{align}
\dot{\phi}^2_{log} &= \rho_{\Lambda,log} \left[ 1 + 3\varepsilon - \frac{M_p^2/3}{3\alpha M_p^2 + \gamma_1 R \log(M_p^2 R^{-1}) + \gamma_2 R} + \frac{1+\Omega_k}{3\Omega_\Lambda} \right], \\
V_{log}(\phi) &= \frac{\rho_{\Lambda,log}}{2} \left[ 1 - 3\varepsilon + \frac{M_p^2/3}{3\alpha M_p^2 + \gamma_1 R \log(M_p^2 R^{-1}) + \gamma_2 R} - \frac{1+\Omega_k}{3\Omega_\Lambda} \right], \\
\phi(a)_{log} - \phi(a_0)_{log} &= \int_{a_0}^{a} \sqrt{3\Omega_{\Lambda,log}} \,
\left[ 1 + 3\varepsilon - \frac{M_p^2/3}{3\alpha M_p^2 + \gamma_1 R \log(M_p^2 R^{-1}) + \gamma_2 R} + \frac{1+\Omega_k}{3\Omega_\Lambda} \right]^{1/2} \frac{da}{a}.
\end{align}

In the limiting case of a flat Dark Dominated Universe, i.e. $\omega_{\Lambda,pl} = \Omega_{\Lambda,log} = 1$, $\Omega_k = \Omega_m = 0$, and $\delta = 0$, the scalar field and potential reduce to
\begin{align}
\dot{\phi}^2(t) &= \frac{108 M_p^2 \alpha^2}{(12\alpha-1)^2} \left( \frac{4}{3} - \frac{1}{9\alpha} + 3\varepsilon \right) \frac{1}{t^2}, \label{murano12} \\
V(t) &= \frac{54 M_p^2 \alpha^2}{(12\alpha-1)^2} \left( \frac{2}{3} + \frac{1}{9\alpha} - 3\varepsilon \right) \frac{1}{t^2}.
\end{align}

Integrating $\dot{\phi}^2$, we obtain
\begin{equation}
\phi(t) = \sqrt{\frac{108 M_p^2 \alpha^2}{(12\alpha-1)^2} \left( \frac{4}{3} - \frac{1}{9\alpha} + 3\varepsilon \right)} \, \ln t + \text{constant}.
\end{equation}

In the limiting case $\varepsilon = 0$:
\begin{align}
\phi(t) &= \sqrt{\frac{12 M_p^2 \alpha}{12\alpha-1}} \, \ln t + \text{constant}, \\
V(t) &= \frac{6 M_p^2 \alpha (6\alpha+1)}{(12\alpha-1)^2} \frac{1}{t^2}.
\end{align}

\subsection{The New Modified Chaplygin Gas (NMCG) Scalar Field Model}
We now consider the New Modified Chaplygin Gas (NMCG) as a model for DE, which has the EoS \cite{newm1}:
\begin{equation}
p_D = B \rho_\Lambda - \frac{K(a)}{\rho_\Lambda^\theta}, \label{murano55}
\end{equation}
where $K(a)$ is a function of the scale factor $a$, $B$ is a positive constant, and $0 \leq \theta \leq 1$.  

If we take $K(a)$ in the form \cite{newm2}:
\begin{equation}
K(a) = - \omega_\Lambda A_1 a^{-3(\omega_\Lambda +1)(\theta+1)},
\end{equation}
Eq. \eqref{murano55} becomes
\begin{equation}
p_D = B \rho_\Lambda + \frac{\omega_\Lambda A_1}{\rho_\Lambda^\theta} a^{-3(\omega_\Lambda +1)(\theta+1)}. \label{murano56}
\end{equation}

The energy density $\rho_\Lambda$ of the NMCG model is given by
\begin{equation}
\rho_\Lambda = \left[ \frac{\omega_\Lambda A_1}{\omega_\Lambda - B} a^{-3(\omega_\Lambda +1)(\theta+1)} + B_1 a^{-3(B+1)(\theta+1)} \right]^{\frac{1}{1+\theta}}, \label{murano57}
\end{equation}
where $B_1$ is a constant of integration.  

From Eq. \eqref{murano57}, we obtain
\begin{equation}
B_1 = a^{3(B+1)(\theta+1)} \left[ \rho_\Lambda^{\theta+1} - \frac{\omega_\Lambda}{\omega_\Lambda - B} A_1 a^{-3(\omega_\Lambda+1)(\theta+1)} \right]. \label{murano58}
\end{equation}

Using the definition of the EoS parameter $\omega_\Lambda$, from Eq. \eqref{murano56} we find
\begin{equation}
A_1 = \frac{\omega_\Lambda - B}{\omega_\Lambda} \, \rho_\Lambda^{\theta+1} \, a^{3(\omega_\Lambda+1)(\theta+1)}. \label{murano59}
\end{equation}

For the R-PLECHDE model, inserting $\omega_{\Lambda,pl}$  gives
\begin{align}
B_{1,pl} &= a^{3(B+1)(\theta+1)} \left[ \rho_{\Lambda,pl}^{\theta+1} - \frac{\omega_{\Lambda,pl}}{\omega_{\Lambda,pl}-B} A_{1,pl} a^{-3(\omega_{\Lambda,pl}+1)(\theta+1)} \right], \label{murano58pl} \\
A_{1,pl} &= \frac{\omega_{\Lambda,pl}-B}{\omega_{\Lambda,pl}} \, \rho_{\Lambda,pl}^{\theta+1} \, a^{3(\omega_{\Lambda,pl}+1)(\theta+1)}. \label{murano59pl}
\end{align}

For the R-LECHDE model, we obtain
\begin{align}
B_{1,log} &= a^{3(B+1)(\theta+1)} \left[ \rho_{\Lambda,log}^{\theta+1} - \frac{\omega_{\Lambda,log}}{\omega_{\Lambda,log}-B} A_{1,log} a^{-3(\omega_{\Lambda,log}+1)(\theta+1)} \right], \\
A_{1,log} &= \frac{\omega_{\Lambda,log}-B}{\omega_{\Lambda,log}} \, \rho_{\Lambda,log}^{\theta+1} \, a^{3(\omega_{\Lambda,log}+1)(\theta+1)}.
\end{align}

In the limiting case of a flat Dark Dominated Universe, the parameters of the NMCG model reduce to:
\begin{align}
B_{1, Dark} &= t^{\frac{18\alpha(B+1)(\theta+1)}{12\alpha - 1}} \Bigg\{ 
\left[\frac{108M_p^2 \alpha^2}{(12\alpha-1)^2} \frac{1}{t^2}\right]^{\theta+1} 
- \frac{\frac{1}{3}-\frac{1}{9\alpha}+3\varepsilon}{\frac{1}{3}-\frac{1}{9\alpha}+3\varepsilon - B} 
A_1 \left(t^{\frac{6\alpha}{12\alpha - 1}}\right)^{-(4-\frac{1}{3\alpha}+9\varepsilon)(\theta+1)} 
\Bigg\}, \label{murano65}\\
A_{1, Dark} &= \frac{\frac{1}{3}-\frac{1}{9\alpha}+3\varepsilon - B}{\frac{1}{3}-\frac{1}{9\alpha}+3\varepsilon} 
\left[ \frac{108 M_p^2 \alpha^2}{(12\alpha-1)^2} \right]^{\theta+1} 
t^{\frac{6\alpha}{12\alpha - 1} \left( 4 - \frac{1}{3\alpha} + 9\varepsilon \right)(\theta+1) - 2(\theta+1)}. \label{murano66}
\end{align}

In the limiting case of $\varepsilon = 0$, we have:
\begin{align}
B_{1, Dark, lim} &= \left\{ \left[\frac{108 M_p^2 \alpha^2}{(12\alpha-1)^2} \right]^{\theta+1} 
- \frac{\frac{1}{3}-\frac{1}{9\alpha}}{\frac{1}{3}-\frac{1}{9\alpha}-B} A_1 \right\} 
t^{\frac{18\alpha(B+1)(\theta+1)}{12\alpha-1} - 2(\theta+1)}, \\
A_{1, Dark, lim} &= \frac{\frac{1}{3}-\frac{1}{9\alpha}-B}{\frac{1}{3}-\frac{1}{9\alpha}} 
\left[ \frac{108 M_p^2 \alpha^2}{(12\alpha-1)^2} \right]^{\theta+1}.
\end{align}

Using the general definition of the DE EoS parameter $\omega_\Lambda$, Eqs.~\eqref{gcg6-1} and \eqref{gcg7-1} can be rewritten as:
\begin{align}
\dot{\phi}^2 &= (1+\omega_\Lambda) \rho_\Lambda, \label{gcg6mura1}\\
V(\phi) &= \frac{1}{2}(1-\omega_\Lambda)\rho_\Lambda. \label{gcg7mura2}
\end{align}

For the R-PLECHDE model, we obtain
\begin{align}
\dot{\phi}^2_{pl} &= \rho_{\Lambda, pl} \left[ 1 + 3\varepsilon - \frac{1}{3(3\alpha - \beta R^{\gamma/2 -1})} + \frac{1+\Omega_k}{3\Omega_\Lambda} \right], \\
V_{pl}(\phi) &= \frac{\rho_{\Lambda, pl}}{2} \left[ 1 - 3\varepsilon + \frac{1}{3(3\alpha - \beta R^{\gamma/2 -1})} - \frac{1+\Omega_k}{3\Omega_\Lambda} \right].
\end{align}

The evolutionary form of the scalar field for the NMCG model is then
\begin{align}
\phi(a)_{pl} - \phi(a_0)_{pl} &= \int_{a_0}^a \sqrt{3 \Omega_{\Lambda, pl}} 
\left[ 1 + 3\varepsilon - \frac{1}{3(3\alpha - \beta R^{\gamma/2 -1})} + \frac{1+\Omega_k}{3\Omega_\Lambda} \right]^{1/2} \frac{da}{a}.
\end{align}

Similarly, for the R-LECHDE model:
\begin{align}
\dot{\phi}^2_{log} &= \rho_{\Lambda, log} \left[ 1 + 3\varepsilon - \frac{M_p^2/3}{3\alpha M_p^2 + \gamma_1 R \log(M_p^2 R^{-1}) + \gamma_2 R} + \frac{1+\Omega_k}{3\Omega_\Lambda} \right], \\
V_{log}(\phi) &= \frac{\rho_{\Lambda, log}}{2} \left[ 1 - 3\varepsilon + \frac{M_p^2/3}{3\alpha M_p^2 + \gamma_1 R \log(M_p^2 R^{-1}) + \gamma_2 R} - \frac{1+\Omega_k}{3\Omega_\Lambda} \right], \\
\phi(a)_{log} - \phi(a_0)_{log} &= \int_{a_0}^a \sqrt{3 \Omega_{\Lambda, log}} 
\left[ 1 + 3\varepsilon - \frac{M_p^2/3}{3\alpha M_p^2 + \gamma_1 R \log(M_p^2 R^{-1}) + \gamma_2 R} + \frac{1+\Omega_k}{3\Omega_\Lambda} \right]^{1/2} \frac{da}{a}.
\end{align}

In the limiting case of a flat Dark Dominated Universe, the scalar field and potential reduce to
\begin{align}
\dot{\phi}^2(t) &= \frac{108 M_p^2 \alpha^2}{(12\alpha-1)^2} \left( \frac{4}{3} - \frac{1}{9\alpha} + 3\varepsilon \right) \frac{1}{t^2}, \\
V(t) &= \frac{54 M_p^2 \alpha^2}{(12\alpha-1)^2} \left( \frac{2}{3} + \frac{1}{9\alpha} - 3\varepsilon \right) \frac{1}{t^2}, \\
\phi(t) &= \sqrt{\frac{108 M_p^2 \alpha^2}{(12\alpha-1)^2} \left( \frac{4}{3} - \frac{1}{9\alpha} + 3\varepsilon \right)} \, \ln t + \text{constant}.
\end{align}

In the further limiting case $\varepsilon = 0$:
\begin{align}
\phi(t) &= \sqrt{\frac{12 M_p^2 \alpha}{12\alpha -1}} \, \cdot \ln t + \text{constant}, \\
V(t) &= \frac{6 M_p^2 \alpha (6\alpha +1)}{(12\alpha -1)^2} \cdot\frac{1}{t^2}.
\end{align}

\subsection{The Yang-Mills (YM) Scalar Field Model}
We now turn our attention to the Yang-Mills (YM) model. Recent studies indicate that the Yang-Mills field \cite{ym1,ym9-2,ym9-3,ym9-4,ym9-5} can serve as a viable candidate for describing the nature of dark energy (DE). There are two main reasons supporting the consideration of the YM field as a source of DE. First, in conventional scalar field models, the connection between the field and particle physics frameworks remains unclear. Second, the field inherently respects the weak energy condition, which cannot be violated.

The YM field under consideration exhibits several interesting properties. It forms a fundamental cornerstone of any particle physics model involving interactions mediated by gauge bosons, allowing for a natural incorporation into a consistent unified theory of particle physics. Additionally, the equation of state (EoS) of the effective Yang-Mills condensate (YMC) differs from that of ordinary matter and scalar fields. In particular, it naturally accommodates states with $-1 < \omega < 0$ as well as $\omega < -1$.

In the effective YMC dark energy model, the Lagrangian of the effective Yang-Mills field, $L_{YMC}$, is given by
\begin{equation}
L_{YMC} = \frac{bF}{2}\left( \ln \left|\frac{F}{\kappa^2}\right| -1 \right),
\label{murano117}
\end{equation}
where $\kappa$ denotes the renormalization scale with dimensions of squared mass, and $F$ acts as the order parameter of the YMC, defined as
\begin{equation}
F = -\frac{1}{2}F_{\mu \nu}^{\alpha} F^{\alpha \mu \nu} = E^2 - B^2.
\label{murano118}
\end{equation}
In the purely electric case ($B = 0$), this reduces to $F = E^2$.

The parameter $b$ represents the Callan-Symanzik coefficient \cite{ym18,ym18-1} and, for an $SU(N)$ gauge group, it is expressed as
\begin{equation}
b = \frac{11 N - 2 N_f}{24 \pi^2},
\label{murano119}
\end{equation}
where $N_f$ denotes the number of quark flavors.

For the gauge group $SU(2)$, we have $b = 2\cdot \frac{11}{24\pi^2}$ when the contribution of fermions is neglected, and $b = 2\cdot \frac{5}{24\pi^2}$ when the number of quark flavors is taken as $N_f = 6$. In the case of $SU(3)$, the effective Lagrangian in Eq. (\ref{murano117}) provides a phenomenological description of asymptotic freedom for quarks inside hadrons \cite{ym21,ym21-1}.  

It should be emphasized that the $SU(2)$ YM field introduced here is intended as a model for cosmic dark energy and should not be directly identified with QCD gluon fields or the weak-electromagnetic unification gauge fields, such as $Z^0$ and $W^{\pm}$. The YMC is characterized by an energy scale $\kappa^{1/2} \sim 10^{-3}\,\mathrm{eV}$, much smaller than that of QCD or the electroweak scale. The form in Eq. (\ref{murano117}) can be interpreted as an effective Lagrangian up to 1-loop quantum corrections \cite{ym21,ym21-1}.  

A classical $SU(N)$ YM Lagrangian is given by
\begin{equation}
L = \frac{1}{2g_0^2} F, \label{murano120}
\end{equation}
where $g_0$ is the bare coupling constant. Including 1-loop quantum corrections, $g_0$ is replaced by the running coupling $g$:
\begin{equation}
g_0^2 \rightarrow g^2 = \frac{4\cdot 12\pi^2}{11 N \ln \left( \frac{k}{k_0^2} \right)} = \frac{2}{b \ln \left( \frac{k}{k_0^2}\right)}, \label{murano121}
\end{equation}
where $k$ is the momentum transfer and $k_0$ the energy scale. In the effective theory, we replace $k^2$ with the field strength $F$ as
\begin{equation}
\ln \left( \frac{k}{k_0^2} \right) \rightarrow 2 \ln \left| \frac{F}{\kappa^2 e} \right| = 2 \ln \left| \frac{F}{\kappa^2} - 1 \right|, \label{murano122}
\end{equation}
which reproduces the Lagrangian in Eq. (\ref{murano117}).  

Some key features of this effective YMC action include Lorentz invariance, gauge invariance, asymptotic freedom, and the correct trace anomaly \cite{ym16}. Its logarithmic dependence on the field strength resembles the Coleman-Weinberg scalar effective potential \cite{ym19} and the Parker-Raval effective gravity Lagrangian \cite{ym20}.  

It is worth noting that the renormalization scale $\kappa$ is the only free parameter in this model. Unlike scalar field DE models, the YMC Lagrangian is entirely determined by 1-loop quantum corrections, leaving no freedom to adjust its functional form.  

From Eq. (\ref{murano117}), the energy density $\rho_y$ and pressure $p_y$ of the YMC can be derived as
\begin{align}
\rho_y &= \frac{\epsilon E^2}{2} + \frac{b E^2}{2}, \label{murano123} \\
p_y &= \frac{\epsilon E^2}{6} - \frac{b E^2}{2}, \label{murano124}
\end{align}
where the dielectric constant $\epsilon$ is given by
\begin{equation}
\epsilon = 2 \frac{\partial L_{eff}}{\partial F} = b \ln \left| \frac{F}{\kappa^2} \right|. \label{murano125}
\end{equation}

Alternatively, $\rho_y$ and $p_y$ can be written as
\begin{eqnarray}
\rho_y &=& \frac{1}{2} b \kappa^2 (y + 1) e^y,  \\
p_y &=& \frac{1}{6} b \kappa^2 (y - 3) e^y, \label{murano126127}
\end{eqnarray}
and 
\begin{eqnarray}
\rho_y &=& \frac{1}{2} (y + 1) b E^2, \\
p_y &=& \frac{1}{6} (y - 3) b E^2, \label{murano128129}
\end{eqnarray}
where
\begin{equation}
y = \frac{\epsilon}{b} = \ln \left| \frac{F}{\kappa^2} \right| = \ln \left| \frac{E^2}{\kappa^2} \right|. \label{defiy}
\end{equation}

From these expressions, the equation of state (EoS) parameter of the YMC is
\begin{equation}
\omega_y = \frac{p_y}{\rho_y} = \frac{y - 3}{3 (y + 1)}. \label{omegay}
\end{equation}
At the critical point $\epsilon = 0$, we have $\omega_y = -1$, corresponding to a de Sitter expansion. Near this point, $\epsilon < 0$ gives $\omega_y < -1$, while $\epsilon > 0$ gives $\omega_y > -1$. Therefore, the YMC model naturally realizes both ranges $-1 < \omega_y < 0$ and $\omega_y < -1$.

The expression of $\omega_y$ given in Eq. (\ref{omegay}) leads to the following relation for $y$:
\begin{equation}
y = - \frac{3\left(\omega_y + 1\right)}{3 \omega_y - 1}. \label{murano130}
\end{equation}
To ensure a positive energy density $\rho_y$ in any physically viable model, we require $y > 1$, which implies
\begin{equation}
F > \frac{\kappa^2}{e} \approx 0.368 \, \kappa^2.
\end{equation}

Before considering a specific cosmological model, it is instructive to study $\omega_y$ as a function of $F$. The YMC model exhibits a radiation-like equation of state with $p_y = \frac{1}{2}\rho_y$ and $\omega_y = \frac{1}{2}$ for large dielectric values, i.e. $\epsilon \gg b$ (which implies $F \gg \kappa^2$). On the other hand, at the critical point $\epsilon = 0$ (i.e. $F = \kappa^2$), the YMC behaves as a cosmological constant, with $\omega_y = -1$ and $p_y = -\rho_y$. In this case, the YMC energy density equals the critical energy density $\rho_y = \frac{1}{2} b \kappa^2$ \cite{ym1}.

This remarkable property of the YMC EoS, transitioning from $\omega_y = \frac{1}{3}$ at high energies ($F \gg \kappa^2$) to $\omega_y = -1$ at low energies ($F = \kappa^2$), allows for the existence of a scaling solution for the DE component \cite{ym10,ym10-1}. Importantly, this transition is smooth because $\omega_y$ is a continuous function of $y$ in the range $(-1, \infty)$.  

Regarding the possibility of crossing $\omega_y = -1$, Eq. (\ref{omegay}) shows that $\omega_y$ depends solely on the condensate strength $F$. In principle, $\omega_y < -1$ can occur for $F < \kappa^2$, and the crossing is smooth. However, when the YMC is incorporated as a DE component in a cosmological model along with other components, the value of $F$ evolves dynamically with time. If the YMC does not decay into matter or radiation, $\omega_y$ approaches $-1$ asymptotically but does not cross it. Conversely, if the YMC decays into matter and/or radiation, $\omega_y$ can cross $-1$ and may settle at an asymptotic value, e.g. $-1.17$, depending on the coupling strength. In this regime, all physical quantities $\rho_y$, $p_y$, and $\omega_y$ remain smooth, avoiding the finite-time singularities present in some scalar field models.

Equating the YMC EoS $\omega_y$ with the EoS parameters of the DE model under consideration, $y$ can be written as
\begin{equation}
y = - \frac{3\left(\omega_{\Lambda} + 1\right)}{3 \omega_{\Lambda} - 1}. \label{murano131}
\end{equation}
For the R-PLECHDE model, we obtain
\begin{equation}
y_{pl} = - \frac{3\left(\omega_{\Lambda,pl} + 1\right)}{3 \omega_{\Lambda,pl} - 1},
\end{equation}
and for the R-LECHDE model:
\begin{equation}
y_{log} = - \frac{3\left(\omega_{\Lambda,log} + 1\right)}{3 \omega_{\Lambda,log} - 1}.
\end{equation}

In the limiting case of a flat, dark-energy-dominated Universe, we get
\begin{equation}
y_{Dark} = - \frac{4 - \frac{1}{3\alpha} + 9 \varepsilon}{- \frac{1}{3\alpha} + 9 \varepsilon}. \label{murano134}
\end{equation}
For $\varepsilon = 0$, this reduces to
\begin{equation}
y_{Dark} = 12 \alpha - 1. \label{murano134b}
\end{equation}
Requiring $y > 1$ then implies
\begin{equation}
\alpha > \frac{1}{12}.
\end{equation}

\subsection{The Non Linear Electro-Dynamics (NLED) Scalar Field Model}
We now consider the last model in our study, namely the Non-Linear Electrodynamics (NLED) model. Recently, nonlinear extensions of Maxwell's theory have been proposed to avoid cosmic singularities. Exact solutions of Einstein's field equations coupled with NLED reveal significant nonlinear effects in strong gravitational and magnetic fields. Moreover, General Relativity (GR) combined with NLED can provide a natural mechanism for primordial inflation.

The Lagrangian density for free Maxwell fields is given by \cite{ele1,ele2}
\begin{equation}
L_M = - \frac{F^{\mu \nu} F_{\mu \nu}}{4 \mu}, \label{murano135}
\end{equation}
where $F^{\mu \nu}$ is the electromagnetic field strength tensor and $\mu$ is the magnetic permeability.  

A generalization of the Maxwell Lagrangian up to second-order terms in the fields can be written as
\begin{equation}
L = -\frac{F}{4 \mu_0} + \omega F^2 + \eta F^{*2}, \label{murano136}
\end{equation}
where $\omega$ and $\eta$ are constants, and
\begin{equation}
F^* = F_{\mu \nu}^* F^{\mu \nu}, \label{murano137}
\end{equation}
with $F_{\mu \nu}^*$ denoting the dual of $F_{\mu \nu}$.  

We consider the particular case in which a homogeneous electric field $E$ in a plasma induces an electric current and rapidly decays. Then, the magnetic contribution dominates, $B^2 \gg E^2$, so that $E^2 \approx 0$ and $F \approx 2 B^2$, making $F$ a function of $B$ only.  

The pressure $p_{NLED}$ and energy density $\rho_{NLED}$ of the NLED field are then given by
\begin{align}
p_{NLED} &= \frac{B^2}{6 \mu} \left( 1 - 40 \mu \omega B^2 \right), \label{murano138} \\
\rho_{NLED} &= \frac{B^2}{2 \mu} \left( 1 - 8 \mu \omega B^2 \right). \label{murano139}
\end{align}

The weak energy condition, $\rho_{NLED} > 0$, is satisfied for
\begin{equation}
B < \frac{1}{2 \sqrt{2 \mu \omega}},
\end{equation}
while the pressure becomes negative for
\begin{equation}
B > \frac{1}{2 \sqrt{10 \mu \omega}}.
\end{equation}
The magnetic field can generate dark energy if the strong energy condition is violated, i.e.,
\begin{equation}
\rho_B + 3 p_B < 0 \quad \Rightarrow \quad B > \frac{1}{2 \sqrt{6 \mu \omega}}.
\end{equation}

The equation of state (EoS) parameter $\omega_{NLED}$ of the Nonlinear Electrodynamics field is given by
\begin{equation}
\omega_{NLED} = \frac{p_{NLED}}{\rho_{NLED}} = \frac{1 - 40 \mu \omega B^2}{3 \left( 1 - 8 \mu \omega B^2 \right)}, \label{murano140}
\end{equation}
which can be inverted to express $B^2$ as
\begin{equation}
B^2 = \frac{1 - 3 \omega_{NLED}}{8 \mu \omega \left( 5 - 3 \omega_{NLED} \right)}. \label{murano141}
\end{equation}

By making a correspondence between the EoS parameter of the NLED model and that of the dark energy model under consideration, we obtain
\begin{equation}
B^2 = \frac{1 - 3 \omega_\Lambda}{8 \mu \omega \left( 5 - 3 \omega_\Lambda \right)}. \label{murano142}
\end{equation}

Using Eq. (\ref{murano142}) with the R-PLECHDE model, we get
\begin{equation}
B^2_{pl} = \frac{1 - 3 \omega_{\Lambda,pl}}{8 \mu \omega \left( 5 - 3 \omega_{\Lambda,pl} \right)}. \label{murano143}
\end{equation}

For the R-LECHDE model, we similarly have
\begin{equation}
B^2_{log} = \frac{1 - 3 \omega_{\Lambda,log}}{8 \mu \omega \left( 5 - 3 \omega_{\Lambda,log} \right)}. \label{murano144}
\end{equation}

In the limiting case of a flat, dark-energy-dominated Universe, we obtain
\begin{equation}
B^2_{Dark} = \frac{\frac{1}{3\alpha} - 9 \varepsilon}{8 \mu \omega \left( 4 + \frac{1}{3\alpha} - 9 \varepsilon \right)}. \label{murano145}
\end{equation}

For $\varepsilon = 0$, this reduces to
\begin{equation}
B^2_{Dark} = \frac{1}{8 \mu \omega \left( 12 \alpha + 1 \right)}, \label{murano146}
\end{equation}
which requires
\begin{equation}
\alpha \neq - \frac{1}{12}.
\end{equation}

\section{Conclusions}

In this paper, we studied the power-law entropy-corrected and the  logarithmic entropy-corrected versions of the holographic dark energy (HDE) model, which aim to shed light on the nature of dark energy within the framework of Quantum Gravity. We considered dark energy (DE) interacting with dark matter (DM) in a non-flat Friedmann–Robertson–Walker (FRW) Universe, using the Ricci scalar $R$ as the infrared (IR) cut-off. Additionally, we took into account dissipative effects arising from the bulk viscosity of cosmic fluids. The inclusion of the correction terms is motivated by Loop Quantum Gravity (LQG), one of the most promising approaches to Quantum Gravity.

Using the energy densities of the two models, we derived the equation of state (EoS) parameter $\omega_\Lambda$, the deceleration parameter $q$, and the evolutionary form of the energy density parameter $\Omega'_\Lambda$ for the interacting PLECRDE and LECRDE models. Furthermore, we established correspondences between these models and several scalar field models, in particular the Generalized Chaplygin Gas (GCG), the Modified Variable Chaplygin Gas (MVCG), the New Modified Chaplygin Gas (NMCG),  the Yang-Mills (YM) and the Non Linear Electro-Dynamics (NLED) scalar field models. These correspondences are significant because they reveal how different dark energy candidates are interrelated.

We also analyzed the limiting case of a flat, dark-energy-dominated Universe, i.e., when $\gamma_1 = \gamma_2 = 0$, $\Omega_\Lambda = 1$, and $\Omega_k =  \Omega_m= 0$ for the LECRDE model, and $\beta = 0$, $\Omega_\Lambda = 1$, and $\Omega_k = \Omega_m= 0$ for the PLECRDE model. In this limit, we found that both the EoS parameter $\omega$ and the deceleration parameter $q$ remain constant.

\end{document}